\documentclass[preprint,5p]{elsarticle}

\usepackage{amsmath,amssymb,bm}
\usepackage{lineno,hyperref}

\journal{Physics Letter B}
\begin{document}

\begin{frontmatter}

\title{Multi-task learning on nuclear masses and separation energies with the kernel ridge regression}

\author[PKU]{X. H. Wu}
\author[PKU]{Y. Y. Lu}
\author[PKU]{P. W. Zhao.\corref{cor1}}
\ead{pwzhao@pku.edu.cn}


\cortext[cor1]{Corresponding Author}
\address[PKU]{State Key Laboratory of Nuclear Physics and Technology, School of Physics, Peking University, Beijing 100871, China}

\begin{abstract}
  A multi-task learning (MTL) framework, called gradient kernel ridge regression, for nuclear masses and separation energies is developed by introducing gradient kernel functions to the kernel ridge regression (KRR) approach.
  By taking the WS4 mass model as an example, the gradient KRR network is trained with the mass model residuals, i.e., deviations between experimental and theoretical values of masses and one-nucleon separation energies, to improve the accuracy of theoretical predictions.
  Significant improvements are achieved by the gradient KRR approach in both the interpolation and the extrapolation predictions of nuclear masses and separation energies.
  This demonstrates the advantage of the present MTL framework that integrates the information of nuclear masses and separation energies and improves the predictions for both of them.
\end{abstract}

\end{frontmatter}


\section{Introduction}

Nuclear mass contains a wealth of structure information~\cite{Lunney2003Rev.Mod.Phys.1021}, which can be used to extract nuclear deformation~\cite{Hager2006Phys.Rev.Lett.42504, Roubin2017Phys.Rev.C14310}, shell structure\cite{Ramirez2012Science1207, Wienholtz2013Nature346}, effective interactions~\cite{AlexBrown1998Phys.Rev.C220, Lalazissis2005Phys.Rev.C24312, Zhao2010Phys.Rev.C54319}, etc.
The proton and neutron separation energies deduced from nuclear masses are key inputs in understanding the origin of heavy elements in the universe \cite{Burbidge1957Rev.Mod.Phys.547} through proton capture process~\cite{Schatz1998Phys.Rep., Arnould2003Phys.Rep.} and neutron capture process~\cite{Kaeppeler2011Rev.Mod.Phys., Cowan2021Rev.Mod.Phys.}.
During the past decades, great progress has been made in mass measurements of atomic nuclei, and about 2500 nuclear masses have been measured to date~\cite{Wang2021Chin.Phys.C030003}.
Nevertheless, the masses of a large number of nuclei remain unknown from experiments and cannot be measured at least in the foreseeable future.
Therefore, theoretical predictions for nuclear masses and separation energies are imperative at the present time.
Global model can be traced back to the Bethe-Weizs\"acker (BW) formula based on the famous liquid drop model (LDM)~\cite{Weizsaecker1935ZeitschriftfurPhysik431}.
Efforts have been made in introducing microscopic corrections to the LDM, which are known as the macroscopic-microscopic models, such as the finite-range droplet model (FRDM)~\cite{Moeller2016Atom.DataNucl.DataTables1} and the Weizs\"acker-Skyrme (WS) model~\cite{Wang2014Phys.Lett.B215}.
The microscopic models based on the nonrelativistic and relativistic density functional theories have also been developed~\cite{Geng2005Prog.Theor.Phys.785, Erler2012Nature, Afanasjev2013Phys.Lett.B, Yang2021Phys.Rev.C, Zhang2022Atom.DataNucl.DataTables}.
The root-mean-square (rms) deviations of nuclear masses and separation energies between theoretical models and the available experimental data~\cite{Wang2021Chin.Phys.C030003} range from about 3 MeV for the BW model~\cite{Kirson2008Nucl.Phys.A} to about 300 keV for the WS ones~\cite{Wang2014Phys.Lett.B215}, which is still not sufficient for accurate studies of exotic nuclear structure and astrophysical nucleosynthesis~\cite{Mumpower2015Phys.Rev.C035807, Jiang2021Astrophys.J.29}.
What's more, for nuclei far away from the experimentally known region, the predictions of different models can be very different.

Recently, machine learning (ML) has been widely used in the field of physics~\cite{Carleo2019Rev.Mod.Phys.45002, Boehnlein2021}.
For nuclear physics, ML applications can be traced back to early 1990s~\cite{Gazula1992Nucl.Phys.A1,Gernoth1993Phys.Lett.B1}, and recently, it has been widely adopted to many aspects of nuclear physics, e.g.,
$\beta$-decays~\cite{Niu2019Phys.Rev.C064307, Wu2021Phys.Rev.C},
fusion reaction cross-sections~\cite{Akkoyun2020Nucl.Instrum.MethodsPhys.Res.Sect.BBeamInteract.Mater.Atoms},
charge radii~\cite{Akkoyun2013J.Phys.GNucl.Part.Phys., Ma2020Phys.Rev.C14304, Wu2020Phys.Rev.C54323},
excited states~\cite{Lasseri2020Phys.Rev.Lett.162502, Wang2021Phys.Rev.Ca, Bai2021Phys.Lett.B},
nuclear landscape~\cite{Neufcourt2019Phys.Rev.Lett., Neufcourt2020Phys.Rev.C, Neufcourt2020Phys.Rev.Ca},
fission yields~\cite{Wang2019Phys.Rev.Lett.122501, Lovell2020J.Phys.GNucl.Part.Phys.a},
variational calculations~\cite{Keeble2020Phys.Lett.B135743, Adams2021Phys.Rev.Lett.},
extrapolations for many-body physics~\cite{Negoita2019Phys.Rev.C, Jiang2019Phys.Rev.C, Yoshida2020Phys.Rev.C, Ismail2021Phys.Rev.C},
nuclear energy density functional~\cite{Wu2022Phys.Rev.C}, etc.
In particular for nuclear masses, many ML approaches have been employed to improve its description, such as the kernel ridge regression (KRR)~\cite{Wu2020Phys.Rev.C051301, Wu2021Phys.Lett.B, Guo2022Symmetry},
the radial basis function (RBF)~\cite{Wang2011Phys.Rev.C51303, Niu2013Phys.Rev.C24325, Niu2016Phys.Rev.C54315, Niu2018Sci.Bull.759, Ma2017Phys.Rev.C24302},
the Bayesian neural network (BNN)~\cite{Utama2016Phys.Rev.C14311, Utama2017Phys.Rev.C44308, Niu2018Phys.Lett.B48, Neufcourt2018Phys.Rev.C34318},
the Levenberg-Marquardt neural network~\cite{Zhang2017J.Phys.GNucl.Part.Phys.},
the gaussian process regression~\cite{Shelley2021Universe5},
the light gradient boosting machine~\cite{Gao2021Nucl.Sci.Tech.109},
the Bayesian probability classifier~\cite{Liu2021Phys.Rev.C}, etc.
By training the ML network with the mass model residuals, i.e., deviations between experimental and calculated masses, ML approaches can significantly reduce the corresponding rms deviation to about $200$ keV~\cite{Wu2020Phys.Rev.C051301, Niu2013Phys.Rev.C24325, Niu2018Phys.Lett.B48, Shelley2021Universe5}, which can be further reduced to below $150$ keV with the odd-even effects being taken into account~\cite{Wu2021Phys.Lett.B, Niu2016Phys.Rev.C54315}.
Among these studies, it is found that the KRR approach can avoid the risk of worsening the mass predictions for nuclei at large extrapolation distances~\cite{Wu2020Phys.Rev.C051301}, and has obtained the most precise machine-learning mass model so far~\cite{Wu2021Phys.Lett.B}.

Multi-task learning (MTL) is a subfield of machine learning, in which multiple related learning tasks are solved at the same time by exploiting commonalities and differences across tasks~\cite{Caruana1997Machinelearning}.
Comparing with the single-task learning, where related tasks are solved separately, the MTL can improve the generalization performances of all the tasks, since useful information contained in multiple related tasks is shared in the learning procedure~\cite{Zhang2017Natl.Sci.Rev.}.
It was successfully applied in nuclear physics to the description of giant dipole resonance key parameters~\cite{Bai2021Phys.Lett.B}.
But the existing ML studies of nuclear masses and separation energies are all based on single-task learning.
At a first glance, one may doubt the necessity of using MTL for nuclear masses and separation energies, since separation energies can be deduced as the derivatives of nuclear masses.
However, it should be noted that the direct information of the derivative of a function can help to better determine the function itself.
Indeed, this idea has been applied in ML studies of density functional theory~\cite{Meyer2020J.Chem.TheoryComput.}, where simultaneous trainings on the functional and its functional derivative are performed and the predictive accuracies of both the functional and the functional derivative are improved.

In the present work, a MTL framework, i.e., the gradient KRR, for nuclear masses and separation energies based on the KRR approach is developed by introducing gradient kernel functions to the KRR approach.
Both the gaussian kernel function and remodulated kernel function with odd-even effects are adopted in the present framework.
The weight parameters contained in the gradient KRR approach are determined by learning the data of nuclear masses and separation energies simultaneously.
The description precision and extrapolation reliability of the gradient KRR approach are analyzed in detail.

\section{Theoretical framework}

The mass residual of the nucleus $(Z_j, N_j)$ in the KRR approach~\cite{Wu2020Phys.Rev.C051301, Wu2021Phys.Lett.B, Guo2022Symmetry} is expressed as
\begin{equation}\label{KRR_function}
  M^{\rm KRR}_{\rm res}(Z_j,N_j) = \sum_{i=1}^{m} K[(Z_i,N_i),(Z_j,N_j)]\alpha_i,
\end{equation}
where $\alpha_i$ are weights to be determined, $m$ is the number of the nuclei in the training set,
and $K[(Z_i,N_i),(Z_j,N_j)]$ is the kernel function.
In the flat version of the KRR approach~\cite{Wu2020Phys.Rev.C051301}, the kernel function is taken as
\begin{equation}\label{krr_kernel}
  K[(Z_i,N_i),(Z_j,N_j)] = \exp\left[ -\frac{(Z_i-Z_j)^2+(N_i-N_j)^2}{2\sigma^2} \right],
\end{equation}
while later, it is updated as remodulated kernel function in the extended KRR approach with odd-even effects (KRRoe)~\cite{Wu2021Phys.Lett.B},
\begin{align}
  &K[(Z_i,N_i),(Z_j,N_j)]  = \exp\left[ -\frac{(Z_i-Z_j)^2+(N_i-N_j)^2}{2\sigma^2} \right] \label{krroe_kernel} \\
  & ~~~~~~~~~~ +\delta_{\rm oe} \exp\left[ -\frac{(Z_i-Z_j)^2+(N_i-N_j)^2}{2\sigma_{\rm oe}^2} \right]\cdot \frac{\lambda}{\lambda_{\rm oe}}, \notag
\end{align}
where $\delta_{\rm oe}=1$ for two nuclei have the same parity of proton and neutron numbers, otherwise $\delta_{\rm oe}=0$. Here, $\sigma$, $\lambda$, $\sigma_{\rm oe}$ and $\lambda_{\rm oe}$ are hyperparameters to be determined.

In the present gradient KRR approach (GKRR), the mass residual of the nucleus $(Z_j, N_j)$ is expressed as
\begin{align}
  M^{\rm GKRR}_{\rm res}(Z_j,N_j) = & \sum_{i=1}^{m} K[(Z_i,N_i),(Z_j,N_j)]\alpha_i \label{KRR_mass} \\
  & + \sum_{i=1}^m \hat{\bm \Delta}^T_{Z_i,N_i}K[(Z_i,N_i),(Z_j,N_j)]{\bm \beta}_i, \notag
\end{align}
where $\alpha_i$ and ${\bm \beta}_i = \begin{bmatrix} \beta_i^Z \\ \beta_i^N \end{bmatrix}$ are weight parameters, and
\begin{align}
  & \hat{\bm \Delta}_{Z_i,N_i}K[(Z_i,N_i),(Z_j,N_j)] \equiv  \begin{bmatrix} \hat{\Delta}_{Z_i} \\ \hat{\Delta}_{N_i} \end{bmatrix}K[(Z_i,N_i),(Z_j,N_j)] \notag \\
  & \equiv \begin{bmatrix} K[(Z_i,N_i),(Z_j,N_j)]-K[(Z_i-1,N_i),(Z_j,N_j)] \\ K[(Z_i,N_i),(Z_j,N_j)] - K[(Z_i,N_i-1),(Z_j,N_j)] \end{bmatrix},  \label{gradient_terms}
\end{align}
is the so-called gradient kernel functions.

The one-proton and one-neutron separation energy residuals of the nucleus $(Z_j, N_j)$ predicted by the gradient KRR approach can be deduced from the gradient of the mass residual~\eqref{KRR_mass},
\begin{align}
  {\bm S}^{\rm GKRR}_{\rm res}(Z_j,N_j) \equiv & \begin{bmatrix} S^{\rm GKRR}_{p,{\rm res}}(Z_j,N_j) \\ S^{\rm GKRR}_{n,{\rm res}}(Z_j,N_j) \end{bmatrix} \label{KRR_separation_energy} \\
  = & \sum_{i=1}^{m} \hat{\bm \Delta}_{Z_j,N_j}K[(Z_i,N_i),(Z_j,N_j)]\alpha_i \notag \\
  +& \sum_{i=1}^m \hat{\bm \Delta}_{Z_j,N_j}\hat{\bm \Delta}^T_{Z_i,N_i}K[(Z_i,N_i),(Z_j,N_j)]{\bm \beta}_i. \notag
\end{align}

The predictions of nuclear masses~\eqref{KRR_mass} and separation energies~\eqref{KRR_separation_energy} can be rewritten in a more compact form,
\begin{equation}\label{GKRR_prediction}
  {\bm D}^{\rm GKRR}_{\rm res} = K^G {\bm \gamma},
\end{equation}
where
\begin{equation}
  {\bm D}^{\rm GKRR}_{\rm res} \equiv {\begin{bmatrix} \vdots \\ M^{\rm GKRR}_{\rm res}(Z_j,N_j) \\ \vdots  \\ \vdots \\ {\bm S}^{\rm GKRR}_{\rm res}(Z_j,N_j) \\  \vdots  \end{bmatrix}}_{3n\times1}~~~~~{\rm and}~~~~~ {\bm \gamma} = {\begin{bmatrix} \vdots \\ \alpha_i \\ \vdots  \\ \vdots \\ {\bm \beta}_i \\  \vdots  \end{bmatrix}}_{3m\times1},
\end{equation}
are column vectors that contain the predictions of nuclear masses and separation energies, and the weight parameters $\alpha_i$ and ${\bm \beta}_i$, respectively.
The subscripts of the vector denote its dimension, and $n$ is the number of the nuclei to be predicted.
$K^G$ is a kernel matrix defined as
\begin{equation}\label{KG}
  K^G_{3n\times 3m} \equiv \begin{bmatrix} K_{n\times m} & K'_{n\times 2m} \\ K''_{2n\times m} & K'''_{2n\times 2m} \end{bmatrix},
\end{equation}
where
\begin{align}
  & K_{n\times m}(i,j)=K[(Z_i,N_i),(Z_j,N_j)], \\
  & K'_{n\times 2m}(i,j) = \hat{\bm \Delta}^T_{Z_i,N_i}K[(Z_i,N_i),(Z_j,N_j)],\\
  & K''_{2n\times m}(i,j) = -\hat{\bm \Delta}_{Z_j,N_j}K[(Z_i,N_i),(Z_j,N_j)],\\
  & K'''_{2n\times 2m}(i,j) = -\hat{\bm \Delta}_{Z_j,N_j}\hat{\bm \Delta}^T_{Z_i,N_i}K[(Z_i,N_i),(Z_j,N_j)].
\end{align}

The weight parameters ${\bm \gamma}$ are determined by minimizing the loss function defined as
\begin{equation}\label{loss}
  L({\bm \gamma})= ({\bm D}^{\rm GKRR}_{\rm res}-{\bm D}^{\rm Data}_{\rm res})^T({\bm D}^{\rm GKRR}_{\rm res}-{\bm D}^{\rm Data}_{\rm res}) + \lambda{\bm \gamma}^T K^G {\bm \gamma},
\end{equation}
where ${\bm D}^{\rm Data}_{\rm res}$ contains the training data of mass residuals and separation energy residuals.
The first term of \eqref{loss} is the variance between the data ${\bm D}^{\rm Data}_{\rm res}$ and the GKRR prediction ${\bm D}^{\rm GKRR}_{\rm res}$, and the second term is a regularizer that penalizes larger weights to reduce the risk of overfitting.
The hyperparameter $\lambda$ determines the regularization strength.
Minimizing Eq.~\eqref{loss} yields
\begin{equation}\label{gamma_weights}
  {\bm \gamma} = (K^G+\lambda)^{-1}{\bm D}^{\rm Data}_{\rm res}.
\end{equation}
Note that the data of both the nuclear masses and separation energies are used to determine the weight parameters ${\bm \gamma}$, and the network with weight parameters ${\bm \gamma}$ is used to predict both the nuclear masses and separation energies.
This matches the idea of MTL, i.e., information contained in related tasks is shared to make predictions for all the tasks.

Once the weight parameters are determined, the mass residuals and separation energy residuals, i.e., the differences between the experimental data $(M^{\rm exp}, {\bm S}^{\rm exp})$ and the predictions for a given theoretical model $(M^{\rm th}, {\bm S}^{\rm th})$ can be predicted by the GKRR predictions \eqref{GKRR_prediction}.
The mass and separation energies for a nucleus $(Z_j,N_j)$ are thus given as $M^{\rm GKRR}(Z_j,N_j) = M^{\rm th}(Z_j,N_j)+M^{\rm GKRR}_{\rm res}(Z_j,N_j)$ and ${\bm S}^{\rm GKRR}(Z_j,N_j) = {\bm S}^{\rm th}(Z_j,N_j)+{\bm S}^{\rm GKRR}_{\rm res}(Z_j,N_j)$, respectively.

\section{Numerical details}

Both kernel functions \eqref{krr_kernel} and \eqref{krroe_kernel} are adopted in the present work, and the corresponding results are denoted as KRR and KRRoe respectively for the KRR results, and GKRR and GKRRoe respectively for the GKRR ones.
Note that the present gradient KRR approach does not involve new hyperparameters, therefore the hyperparameters $(\sigma=2.38,\lambda=0.3)$ optimized in the KRR study~\cite{Wu2020Phys.Rev.C051301}, and the hyperparameters $(\sigma=1.25,\lambda=0.05,\sigma_{\rm oe}=2.65,\lambda_{\rm oe}=0.15)$ optimized in the KRRoe study~\cite{Wu2021Phys.Lett.B}, are adopted in the present work.

Moreover, the experimental data of 2021 nuclei with $Z\geq8$ and $N\geq8$ are taken from AME2020~\cite{Wang2021Chin.Phys.C030003}.
Note that only the nuclei which have data of nuclear masses, one-proton and one-neutron separation energies are included.
The theoretical data are taken from the mass model WS4~\cite{Wang2014Phys.Lett.B215}.

\section{Results and discussion}

\begin{figure}[!htbp]
  \centering
  \includegraphics[width=0.45\textwidth]{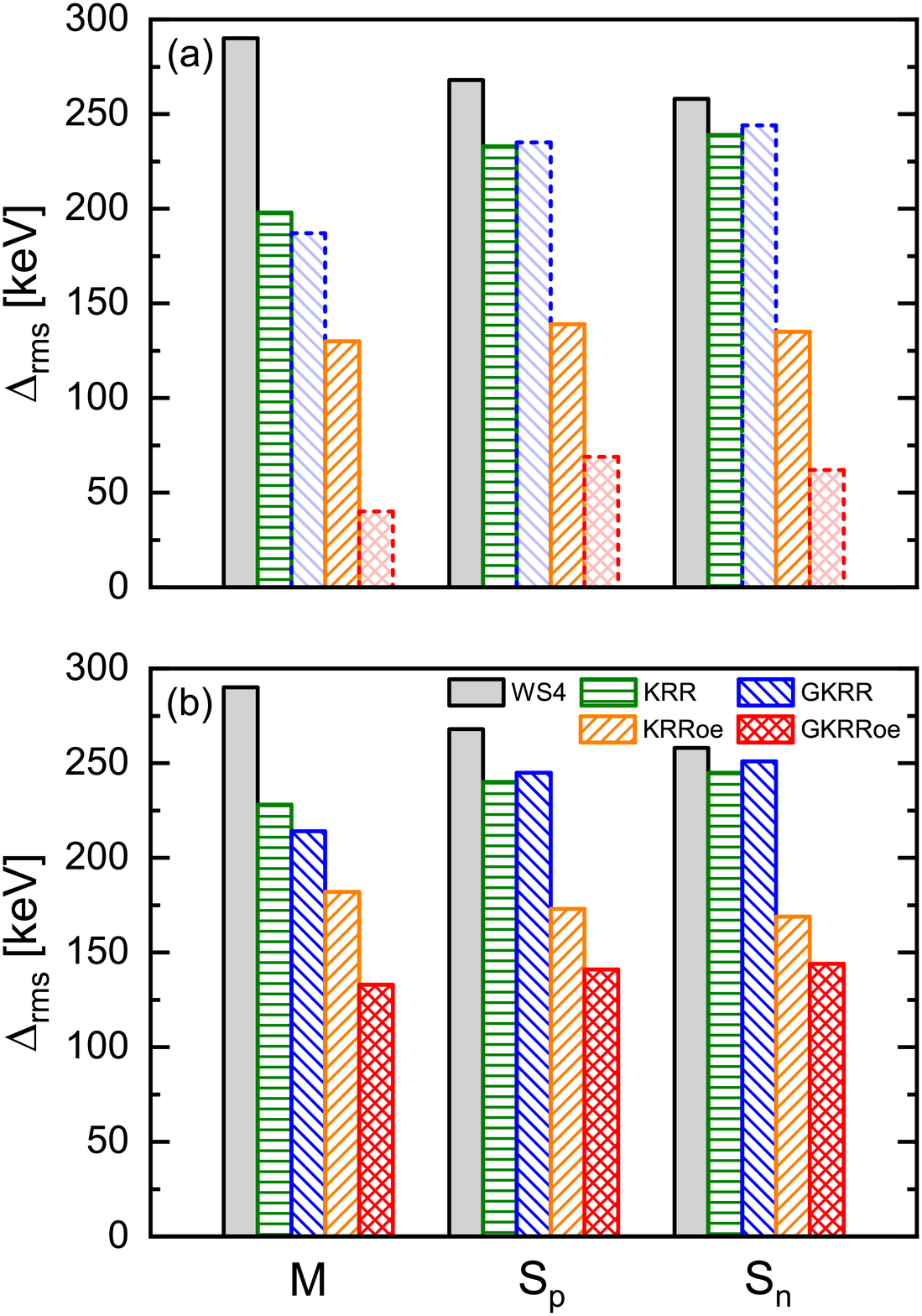}
  \caption{The rms deviations $\Delta_{\rm rms}$ of the predicted masses $M$, one-proton separation energies $S_p$, and one-neutron separation energies $S_n$ from the WS4 mass model, the KRR, GKRR, KRRoe, and GKRRoe approaches with respect to the experimental values in the leave-one-out cross-validation (a) and leave-one-plus-four-out cross-validation (b). The GKRR and GKRRoe results shown in panel (a) are denoted by dashed bars since they should not be regarded as the description precisions (see text).
  }\label{fig1}
\end{figure}

Firstly, the leave-one-out cross-validation (LOOCV) is adopted to evaluate the accuracies of the KRR, GKRR, KRRoe, and GKRRoe predictions.
In each case, the mass and separation energies for a given nucleus are predicted with the ML network trained on all other 2020 nuclei. In Fig.~\ref{fig1}(a), the rms deviations $\Delta_{\rm rms}$ of the predicted masses $M$, one-proton separation energies $S_p$ and one-neutron separation energies $S_n$ from the WS4 mass model, the KRR, GKRR, KRRoe, and GKRRoe results with respect to the experimental values are depicted.
The KRR and GKRR give quite similar results, while with the odd-even effects, the GKRRoe gives better results than the KRRoe.
This indicates that once we include the one-nucleon separation energies in the training set, a good design for the ML model including the odd-even effects is also crucial to achieve a good performance.
As seen in Fig.~\ref{fig1}(a), the GKRRoe results provide the best description of the mass and one-nucleon separation energies, and the rms deviation $\Delta_{\rm rms}$ is around $50$ keV.
However, it should be noted that the LOOCV recipe is not so compatible with the GKRR framework, because the target values can be exactly deduced from the data involved in the training set.
For instance, the mass and separation energies of a given nucleus $(Z,N)$ can be directly deduced from the mass and separation energies data of its four neighboring nuclei $(Z+1,N)$, $(Z,N+1)$, $(Z-1,N)$ and $(Z,N-1)$.
Therefore, in Fig.~\ref{fig1}(a), the GKRR and GKRRoe results are denoted by dashed bars.
If the hyperparameters $\sigma$ and $\sigma_{\rm oe}$ are adjusted to very small values so that the Gaussian kernel functions can only connect the target nucleus with its neighboring nuclei, the deviations of the GKRRoe in the LOOCV can even reach to zero.
However, in that case, the GKRRoe would totally lose its extrapolation power.

In order to evaluate the predictive power of the GKRR and GKRRoe properly, we introduce leave-one-plus-four-out cross-validation (LOFOCV) in the present work.
In LOFOCV, to predict the mass and separation energies for a given nucleus, not only the nucleus itself, but also its four neighboring nuclei are excluded out of the training set.
In such a way, the mass and separation energies of the target nucleus cannot be directly deduced from the training data and, thus, can be regarded as predictions.
For a fair comparison, the KRR, GKRR, KRRoe, and GKRRoe results involved in the present work are all given by the LOFOCV, and the corresponding rms deviations $\Delta_{\rm rms}$ are shown in Fig.~\ref{fig1}(b).

As can be seen in Fig.~\ref{fig1}(b), all ML results significantly improve the descriptions of nuclear masses, but only the KRRoe and GKRRoe results can obviously improve the descriptions of one-nucleon separation energies.
This is understandable, since the one-nucleon separation energies have obvious odd-even staggering behaviors, which can be properly described only when the odd-even effects are taken into account.
The GKRRoe provides the best descriptions for nuclear masses and separation energies.
This indicates the success of the present MTL framework that integrates the information from nuclear masses and separation energies and improves the predictions for both of them.

\begin{figure}[!htbp]
  \centering
  \includegraphics[width=0.45\textwidth]{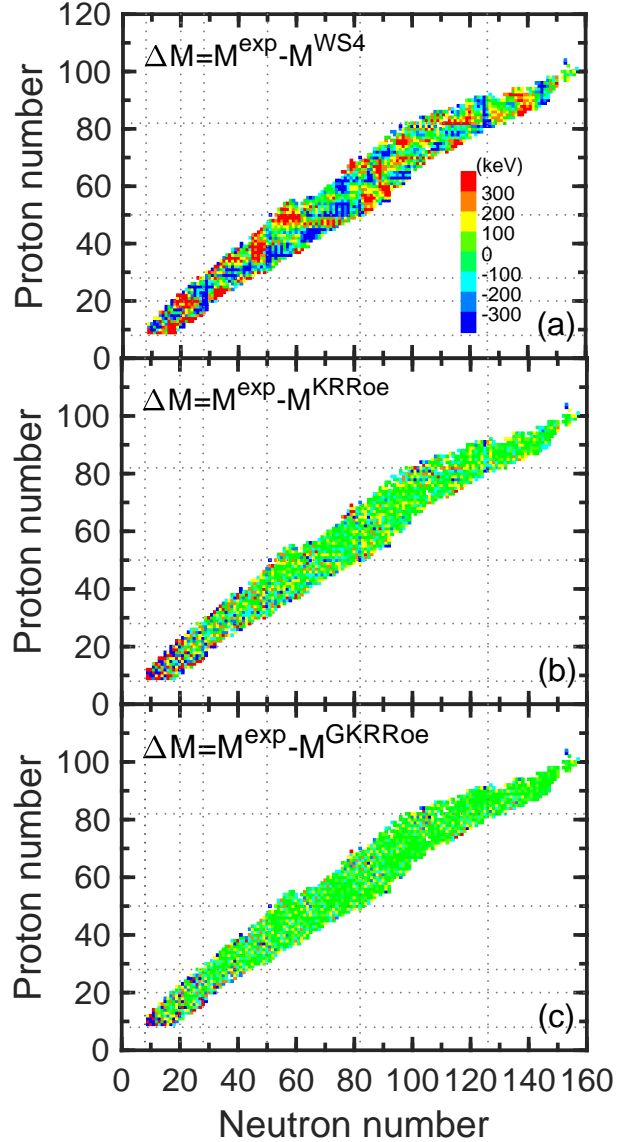}
  \caption{Mass differences between the experimental data~\cite{Wang2021Chin.Phys.C030003} and the results of WS4 (a), KRRoe (b), and GKRRoe (c).
  The magic numbers are indicated with the dotted lines.
  }\label{fig2}
\end{figure}

Mass differences between the experimental data~\cite{Wang2021Chin.Phys.C030003} and the results of WS4, KRRoe, and GKRRoe for 2021 nuclei are shown in Fig.~\ref{fig2}.
The differences between the experimental data and the WS4 mass model are roughly $300$ keV for most nuclei, and this gives rise to a rms deviation of $290$ keV.
This description is greatly improved with the KRRoe results [see Fig.~\ref{fig2}(b)].
The corresponding differences are mainly within $200$ keV, and the corresponding rms deviation is reduced to $182$ keV.
As seen in Fig.~\ref{fig2}(c), the mass description can be further improved by the GKRRoe results, in particular for the medium-mass and heavy nuclei.
Visible differences are mostly seen in the light nuclei region.

\begin{figure}[!htbp]
  \centering
  \includegraphics[width=0.45\textwidth]{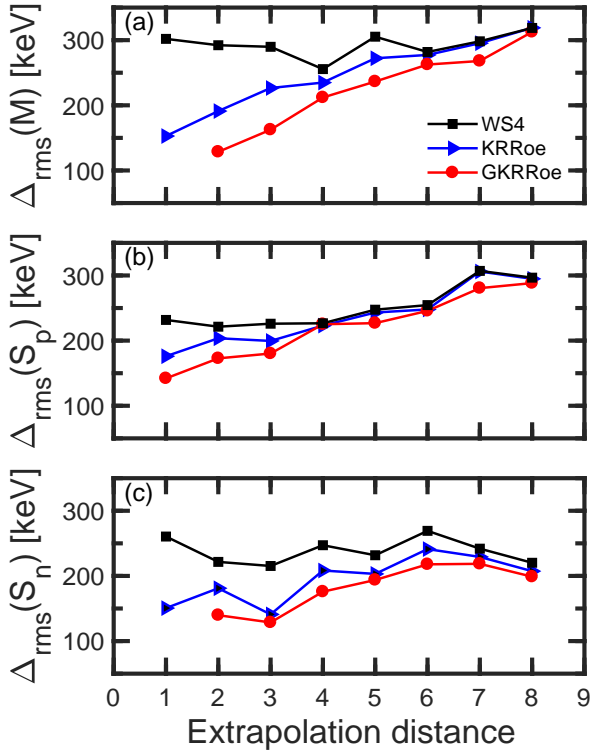}
  \caption{Comparison of the extrapolation power of the KRRoe and the GKRRoe approaches in predicting masses (a), one-proton (b) and one-neutron (c) separation energies for the eight test sets with different extrapolation distances (see text for details).
  }\label{fig3}
\end{figure}

To examine the extrapolation power of the GKRRoe approach, similar to Refs.~\cite{Wu2020Phys.Rev.C051301, Wu2021Phys.Lett.B}, for each isotopic chain, the eight most neutron-rich nuclei are removed out from the training set, and then, they are classified into eight test sets according to their extrapolation distances from the remain training set in the neutron direction.
In Fig.~\ref{fig3}, the rms deviations $\Delta_{\rm rms}$ of the calculated masses, one-proton, and one-neutron separation energies for the eight test sets from the WS4 mass model, the KRRoe, and the GKRRoe extrapolations with respect to the experimental data are shown as functions of the extrapolation distance.

For both the KRRoe and GKRRoe approaches, the rms deviations of masses, one-proton and one-neutron separation energies increase gradually with the extrapolation distance but are still similar or smaller than the WS4 ones at large extrapolation distances.
This is a general feature of the KRR-based approaches with the Gaussian kernel.
As discussed in Refs.~\cite{Wu2020Phys.Rev.C051301, Wu2021Phys.Lett.B}, the KRR reconstructed functions decrease smoothly with the increasing extrapolation distance due to the adoption of Gaussian kernel.
This actually gives the KRR-based approaches the ability to avoid the risk of worsening the predictions for nuclei at large extrapolation distances.

The KRRoe improves the mass predictions of nuclei with the extrapolation distance smaller than 6, and the improvement is up to $100$ keV for small extrapolation distances.
The GKRRoe performs globally better than the KRRoe, not only in the length of the extrapolation distance, but also in the reduction of the $\Delta_{\rm rms}$ at a given distance.
Similar conclusions hold also for the one-nucleon separation energies.
Quantitatively, the improvements for the one-proton separation energies $S_{p}$ are not as large as that for the neutron ones.
This is due to the fact that the neutron excess of the nucleus $(Z-1,N)$ is larger than that of the nucleus $(Z,N-1)$ in most cases.

Therefore, one could conclude that the GKRRoe approach provides significant improvements in the extrapolation predictions of nuclear masses, one-proton, and one-neutron separation energies.
This indicates that such a MTL framework that integrates the information of nuclear masses and separation energies, can help to improve not only the description accuracy but also the extrapolation power.

\section{Summary}

In summary, a MTL framework, called gradient KRR, for nuclear masses and separation energies is developed by introducing gradient kernel functions to the KRR approach.
The newly developed gradient KRR approach is then extended by taking into account the odd-even effects, i.e., the GKRRoe approach.
By taking the WS4 mass model as an example, the GKRRoe network is trained with the mass model residuals, i.e., deviations between experimental and theoretical values of masses and one-nucleon separation energies, to improve the accuracy of theoretical predictions.
Significant improvements in predicting nuclear masses and separation energies for experimentally known nuclei are achieved by the GKRRoe approach.
Moreover, it is found that the GKRRoe approach performs well in extrapolation predictions, not only in the length of the effective extrapolation distance, but also in the reduction of the rms deviation at a given distance.
These results demonstrate the advantage of the present MTL framework that integrates the information of nuclear masses and separation energies and improves the predictions for both of them.


\section*{Acknowledgments}

This work was partly supported by the National Key R\&D Program of China (Contracts No. 2018YFA0404400 and No. 2017YFE0116700), the National Natural Science Foundation of China (Grants No. 11875075, No. 11935003, No. 11975031, No. 12141501 and No. 12070131001), the China Postdoctoral Science Foundation under Grant No. 2021M700256, and the High-performance Computing Platform of Peking University.

\section*{References}

\bibliography{paper}

\end{document}